\documentclass{elsart}
\usepackage{graphicx}
\usepackage[usenames]{color}

\def\ifm#1{\relax\ifmmode#1\else$#1$\fi}

\def\f{\ifm{\phi}}    

\def\ab{\ifm{\sim}}  
\def\x{\ifm{\times}}

\def\pt#1,#2,{\ifm{#1\x10^{#2}}}   
\def\order#1,{\ifm{\mathcal{O}(10^{#1})}}
\def\epm{\ifm{e^+e^-}}
\renewcommand{\to}{\ensuremath{\rightarrow}}

\makeatletter
\newdimen\z@ \z@=0pt 
\newskip\z@skip \z@skip=0pt plus0pt minus0pt
\def\m@th{\mathsurround=\z@}
\def\ialign{\everycr{}\tabskip\z@skip\halign} 
\def\eqalign#1{\null\,\vcenter{\openup\jot\m@th              
  \ialign{\strut\hfil$\displaystyle{##}$&$\displaystyle{{}##}$\hfil
      \crcr#1\crcr}}\,}
\makeatother

\newcommand{\ep}{\mbox{$e^+$}}
\newcommand{\el}{\mbox{$e^-$}}
\newcommand{\ks}{\mbox{$K_S$}}
\newcommand{\kl}{\mbox{$K_L$}}

\newcommand{\pip}{\mbox{$\pi^+$}}
\newcommand{\pim}{\mbox{$\pi^-$}}
\newcommand{\pio}{\mbox{$\pi^{0}$}}

\newcommand{\ppg}{\mbox{$\pi\pi\gamma$}}

\newcommand{\mpp}{\mbox{$M_{\pi\pi}$}}
\newcommand{\mpg}{\mbox{$M_{\pi\gamma}$}}

\newcommand{\dafne}{\mbox{DA$\Phi$NE}}
\newcommand{\br}{\mbox{BR}}

\newcommand{\lint}{\mbox{$L_{\rm int}$}}
\newcommand{\0}{\phantom{0}}
\newcommand{\wpo}{\mbox{$\omega\pio$}}
\newcommand{\wpc}{\mbox{$\pip\pim\pio\pio$}}
\newcommand{\wpn}{\mbox{$\pio\pio\gamma$}}
\newcommand{\eeto}{\mbox{$\ep\el\to\,\,$}}
\newcommand{\sqrts}{\mbox{$\sqrt{s}$}}

\newcommand{\affuni}[2]{Dipartimento di Fisica dell'Universit\`a #1, #2, Italy.}

\newcommand{\affinfnm}[2]{INFN Sezione di #2, #2, Italy.}
\newcommand{\affinfnn}[2]{INFN Sezione di #1, #2, Italy.}

\newcommand{\AmS}{{\protect\the\textfont2
  A\kern-.1667em\lower.5ex\hbox{M}\kern-.125emS}}

\begin{document}

\begin{frontmatter}

\title{\boldmath Study of the process \eeto\wpo\ in the \f-meson 
  mass region with the KLOE detector}

\collab{The KLOE Collaboration}

\author[Na,infnNa]{F.~Ambrosino},
\author[Frascati]{A.~Antonelli},
\author[Frascati]{M.~Antonelli},
\author[Roma2,infnRoma2]{F.~Archilli},
\author[Karlsruhe]{P.~Beltrame},
\author[Frascati]{G.~Bencivenni},
\author[Frascati]{S.~Bertolucci},
\author[Roma1,infnRoma1]{C.~Bini},
\author[Frascati]{C.~Bloise},
\author[Roma3,infnRoma3]{S.~Bocchetta},
\author[Frascati]{F.~Bossi},
\author[infnRoma3]{P.~Branchini},
\author[Frascati]{P.~Campana},
\author[Frascati]{G.~Capon},
\author[Frascati]{T.~Capussela},
\author[Roma3,infnRoma3]{F.~Ceradini},
\author[Frascati]{P.~Ciambrone},
\author[Roma1]{F.~Crucianelli},
\author[Frascati]{E.~De~Lucia},
\author[Roma1,infnRoma1]{A.~De~Santis\corauthref{cor}},
\ead{antonio.desantis@roma1.infn.it}
\corauth[cor]{Corresponding author.}
\author[Frascati]{P.~De~Simone},
\author[Roma1,infnRoma1]{G.~De~Zorzi},
\author[Karlsruhe]{A.~Denig},
\author[Roma1,infnRoma1]{A.~Di~Domenico},
\author[infnNa]{C.~Di~Donato},
\author[Roma3,infnRoma3]{B.~Di~Micco},
\author[Frascati]{M.~Dreucci},
\author[Frascati]{G.~Felici},
\author[Frascati]{M.~L.~Ferrer},
\author[Roma1,infnRoma1]{S.~Fiore},
\author[Roma1,infnRoma1]{P.~Franzini},
\author[Frascati]{C.~Gatti},
\author[Roma1,infnRoma1]{P.~Gauzzi},
\author[Frascati]{S.~Giovannella\corauthref{cor}}
\ead{simona.giovannella@lnf.infn.it}
\author[infnRoma3]{E.~Graziani},
\author[Karlsruhe]{W.~Kluge},
\author[Frascati]{G.~Lanfranchi},
\author[Frascati,StonyBrook]{J.~Lee-Franzini},
\author[Karlsruhe]{D.~Leone},
\author[Frascati,Energ]{M.~Martini},
\author[Na,infnNa]{P.~Massarotti},
\author[Na,infnNa]{S.~Meola},
\author[Frascati]{S.~Miscetti},
\author[Frascati]{M.~Moulson},
\author[Frascati]{S.~M\"uller},
\author[Frascati]{F.~Murtas},
\author[Na,infnNa]{M.~Napolitano},
\author[Roma3,infnRoma3]{F.~Nguyen},
\author[Frascati]{M.~Palutan},
\author[infnRoma1]{E.~Pasqualucci},
\author[infnRoma3]{A.~Passeri},
\author[Frascati,Energ]{V.~Patera},
\author[Na,infnNa]{F.~Perfetto},
\author[Frascati]{P.~Santangelo},
\author[Frascati]{B.~Sciascia},
\author[Frascati,Energ]{A.~Sciubba},
\author[Frascati]{A.~Sibidanov},
\author[Frascati]{T.~Spadaro},
\author[Roma1,infnRoma1]{M.~Testa},
\author[infnRoma3]{L.~Tortora},
\author[infnRoma1]{P.~Valente},
\author[Frascati]{G.~Venanzoni},
\author[Frascati,Energ]{R.Versaci},
\author[Frascati,Beijing]{G.~Xu}
\address[Frascati]{Laboratori Nazionali di Frascati dell'INFN, 
Frascati, Italy.}
\address[Karlsruhe]{Institut f\"ur Experimentelle Kernphysik, 
Universit\"at Karlsruhe, Germany.}
\address[Na]{Dipartimento di Scienze Fisiche dell'Universit\`a 
``Federico II'', Napoli, Italy}
\address[infnNa]{INFN Sezione di Napoli, Napoli, Italy}
\address[Energ]{Dipartimento di Energetica dell'Universit\`a 
``La Sapienza'', Roma, Italy.}
\address[Roma1]{\affuni{``La Sapienza''}{Roma}}
\address[infnRoma1]{\affinfnm{``La Sapienza''}{Roma}}
\address[Roma2]{\affuni{``Tor Vergata''}{Roma}}
\address[infnRoma2]{\affinfnn{Roma Tor Vergata}{Roma}}
\address[Roma3]{\affuni{``Roma Tre''}{Roma}}
\address[infnRoma3]{\affinfnn{Roma Tre}{Roma}}
\address[StonyBrook]{Physics Department, State University of New 
York at Stony Brook, USA.}
\address[Beijing]{Institute of High Energy 
Physics of Academica Sinica,  Beijing, China.}

%
\begin{abstract}
%
We have studied the \eeto\wpo\ cross section in the \sqrts\ interval 
1000-1030 MeV using the \wpc\ and \wpn\ final 
states with a sample of $\sim$ 600 pb$^{-1}$ collected with the KLOE 
detector at \dafne.
By fitting the observed interference pattern around $M_\phi$ for both 
final states, we extract the ratio of the decay widths
$\Gamma(\omega\to\pio\gamma)/\Gamma(\omega\to\pip\pim\pio) = 
0.0897\pm 0.0016$
and derive the branching fractions 
$\br(\omega\to\pip\pim\pio)= (90.24\pm 0.19)\%$,
$\br(\omega\to\pio\gamma) = (8.09\pm 0.14)\%$.
The parameters describing the \eeto\wpc\ reaction around $M_\phi$ 
are also used to extract the branching fraction for the OZI and G-parity 
violating $\f\to\wpo$ decay: $\br(\f\to\wpo) = (4.4\pm 0.6)\times 10^{-5}$. 
\end{abstract}

\end{frontmatter}

\section{Introduction}

At low energy, below 1.4 GeV, the \eeto\wpo\ cross section is 
largely dominated by the non-resonant processes 
$\eeto \rho/\rho' \to \omega \pio$. 
However, in the region around $M_\phi$, a contribution from the OZI and 
G-parity violating decay \f\to\wpo\ is expected. 
This strongly suppressed decay (\order-5,) can be observed via
interference with the non-resonant process, showing up as a dip 
in the total cross section dependence from \sqrts. \\
The \eeto\wpo\ cross section as a function of \sqrts\ is parametrized
as \cite{phiwp2}:
\begin{equation} \label{eq:xsec}
\sigma(\sqrts) = \sigma_{nr}(\sqrts)
\cdot\left|1-Z\frac{M_{\phi}\Gamma_{\phi}}{D_{\phi}(\sqrts)}\right|^2
\end{equation}
where $\sigma_{nr}(\sqrts)$ is the bare cross section for 
the non-resonant process, $Z$ is the interference parameter 
(i.e. the ratio between the $\phi$ decay and the 
non-resonant process amplitudes), while $M_{\phi}$, $\Gamma_{\phi}$ and $D_{\phi}$
are the mass, the width and the inverse propagator of the $\phi$ meson 
respectively.
The non-resonant cross section, in the energy range of interest, 
increases linearly with \sqrts. A model independent parametrization 
is used to describe the non resonant part:
$\sigma_{nr}(\sqrts) = \sigma_{0} + \sigma' (\sqrts - M_\phi)$.

In this work we study two different final states: \wpc\ and \wpn,
corresponding to the $\omega$ decay in $\pip\pim\pio$ 
ad $\pio\gamma$, respectively. 
From the \wpc\ analysis we have measured the Z parameter that has 
been used to determine the \f\to\wpo\ branching ratio (BR).
In the case of \wpn\ we expect contributions also from 
$\phi\to\rho\pi$ and $\phi\to S\gamma$ intermediate states, being
$S$ a scalar meson.
In \cite{Dalitz_piopiog} we have shown that at $\sqrts \sim M_{\phi}$ 
the interference between $\phi\to S\gamma$ and $\eeto\omega\pio$ events, 
evaluated by fitting the \mpp-\mpg\ Dalitz plot, is small.
Therefore a fit to the 
cross section interference pattern for \wpn\ final state will provide 
information about the $\eeto\rho/\rho'\to\omega\pio$ process and the 
resonant decays \f\to\wpo\ and $\phi\to\rho\pio$. 
Moreover, combining the results we extract the ratio 
$\Gamma(\omega\to\pio\gamma)/\Gamma(\omega\to\pip\pim\pio)$.

\section{The KLOE detector}

The KLOE 
experiment operates at \dafne\ \cite{DAFNE}, the 
Frascati $\phi$-factory. DA$\Phi$NE is an $e^+e^-$ collider running at 
a center of mass energy of $\sim 1020$~MeV, the mass of the 
$\phi$ meson. Equal energy positron and electron beams collide at an 
angle of $\pi$-25 mrad, producing $\phi$ mesons nearly at rest.

The KLOE detector consists of a large cylindrical drift chamber, DC, 
surrounded by a lead-scintillating fiber electromagnetic calorimeter, EMC. 
A superconducting coil around the EMC provides a 0.52 T field. 
The drift chamber~\cite{DCH}, 4~m in diameter and 3.3~m long, has 12,582 
all-stereo tungsten sense wires and 37,746 aluminum field wires. The chamber 
shell is made of carbon fiber-epoxy composite and the gas used is a 90\% 
helium, 10\% isobutane mixture. These features maximize transparency to 
photons and reduce $\kl\to\ks$ regeneration and multiple scattering. The 
position resolutions are $\sigma_{xy}$\ab150 $\mu$m and $\sigma_z$\ab~2 mm. 
The momentum resolution is $\sigma(p_{\perp})/p_{\perp}\approx 0.4\%$. 
Vertices are reconstructed with a spatial resolution of \ab3~mm. 
The calorimeter~\cite{EMC} is
divided into a barrel and two endcaps, for a total of 88 modules, and covers 
98\% of the solid angle. The modules are read out at both ends by 
photo-multipliers, both in amplitude and time. The readout granularity is 
\ab\,(4.4\x4.4)~cm$^2$, for a total of 2440 cells arranged in five layers.  
The energy deposits are obtained from the signal amplitude while the 
arrival times and the particles positions are 
obtained from the time differences.
Cells close in time and space are grouped into calorimeter 
clusters. The cluster energy $E$ is the sum of the cell energies. 
The cluster time $T$ and position $\vec{R}$ 
are energy-weighed averages. Energy and time resolutions are $\sigma_E/E = 
5.7\%/\sqrt{E\ {\rm(GeV)}}$ and  $\sigma_t = 57\ {\rm ps}/\sqrt{E\ {\rm(GeV)}}
\oplus100\ {\rm ps}$, respectively.
The KLOE trigger \cite{TRG} uses both calorimeter and chamber information. 
In this analysis the events are selected by the calorimeter trigger, 
requiring two energy deposits with $E>50$ MeV for the barrel and $E>150$ MeV 
for the endcaps. A cosmic veto rejects events with at least two energy 
deposits above 30 MeV in the outermost calorimeter layer.
Data are then analyzed by an event classification filter 
\cite{NIMOffline}, which streams various categories of events in different 
output files.

\section{Data analysis}
\label{Par:Analysis}

All available statistics collected at the $\phi$ peak during 2001--2002 
data-taking periods, corresponding to 450 pb$^{-1}$, have been analyzed.
Moreover four scan points (at 1010, 1018, 1023 and 1030 MeV) of 
$\sim$10 pb$^{-1}$ each and the off-peak run at 1000 MeV of
$\sim$100 pb$^{-1}$ acquired
in 2005-2006 have been included in this analysis. 
The luminosity is measured with 0.5\% absolute precision counting 
large angle Bhabha scattering events \cite{LUMI}.
Data taken at the $\phi$ peak are grouped in center of mass energy bins 
of 100 keV width. For all the other points, close \sqrts\ values are 
grouped together and the average center of mass energy is evaluated by
weighting with luminosity. 

  In the following, the visible cross section 
  for a given final state (\emph{j}=$4\pi$,\ppg) 
  is defined $\sigma_{\rm vis}^j = N_j/\varepsilon_j\lint$, 
  where $N_j$, $\varepsilon_j$ and \lint, 
  are the number of signal events, 
  the analysis efficiency and the luminosity, respectively.\\
  The visible cross section is related to the bare cross section 
  through the radiator function H (see Sec.~\ref{Par:FitResults}): 
  $$
  \sigma_{\rm vis}^j(\sqrts) = \int ds' H(s,s')\sigma^j(s') = 
  \sigma^j(\sqrts)*\delta_{rad}^j(\sqrts).
  $$

\subsection{$\eeto\omega\pio\to\wpc$}
In the \wpc\ analysis, data are filtered by selecting events with 
the expected final state signature: two tracks with opposite curvature 
connected to a vertex inside a small 
cylindrical fiducial volume ($\rho < 4$ cm and $|z|<6$ cm) around 
the Interaction Point (IP) and four 
neutral clusters in the prompt Time Window (TW), defined as
$|T_{\gamma}-R_{\gamma}/c|<{\rm MIN}(4\,\sigma_t,2\mbox{ ns})$.
To minimize contamination from machine background, we require for the 
clusters a minimal energy of 10 MeV and an angle
with respect to the beam line in the interval 
$22^{\circ} \leq \theta_\gamma \leq 158^{\circ}$.  

With this selection, at $\phi$ resonance, the main background 
contributions come from $\phi\to\ks\kl\to\wpc/\pio\pio\pi\mu\nu$ 
and $\phi\to K^+K^-$, with 
$K^\pm\to\pi^\pm\pio$, which have the same final state signature.
Other two resonant background components ($\phi\to\eta\gamma$ with 
$\eta\to\pip\pim\pio$, and $\phi\to\pip\pim\pio$) mimic the final state 
signature because of additional clusters due to accidental coincidence of
machine background events and/or shower fragments (cluster splitting).
An additional non-resonant background contribution of the order of few 
percent from $\epm\to\pip\pim\pio\pio$, dominated by the $a_1(1260)\pi$ 
intermediate state, is also expected for all \sqrts\ values
\cite{CMD2_a1pi}.

A global kinematic fit ($N_{\rm \! dof}=8$), imposing total four-momentum 
conservation and proper time of flight (TOF) for photons coming from 
the charged vertex, improves both the signal/background separation and the 
determination of the photon energies.
The resulting $\chi^2$ ($\chi^2_{\rm Kfit}$) is used to select  
signal enriched ($\chi^2_{\rm Kfit}<50$), S$_{\rm evts}$, and  
background dominated ($\chi^2_{\rm Kfit}>50$), B$_{\rm evts}$, samples.
In S$_{\rm evts}$ sample the overall contamination from resonant 
background at $\sqrts\sim M_\phi$ is about $12\%$ while becomes negligible 
outside $\phi$ resonance. The contribution of the $a_1(1260)\pi$
background is about 4\% for all \sqrts\ values.
The signal selection efficiency in the S$_{\rm evts}$ sample is
evaluated by Monte Carlo (MC) and corrected with data/MC ratios for 
tracking, vertexing and clustering. The resulting value 
$\varepsilon_{4\pi} \sim 38\%$ is dominated by the selection requirements and 
shows a small dependence on \sqrts, which is taken into account in the 
evaluation of $\sigma_{\rm vis}^{4\pi}$. 

\begin{figure}[ht]
\begin{center}
\includegraphics[width=.9\textwidth]{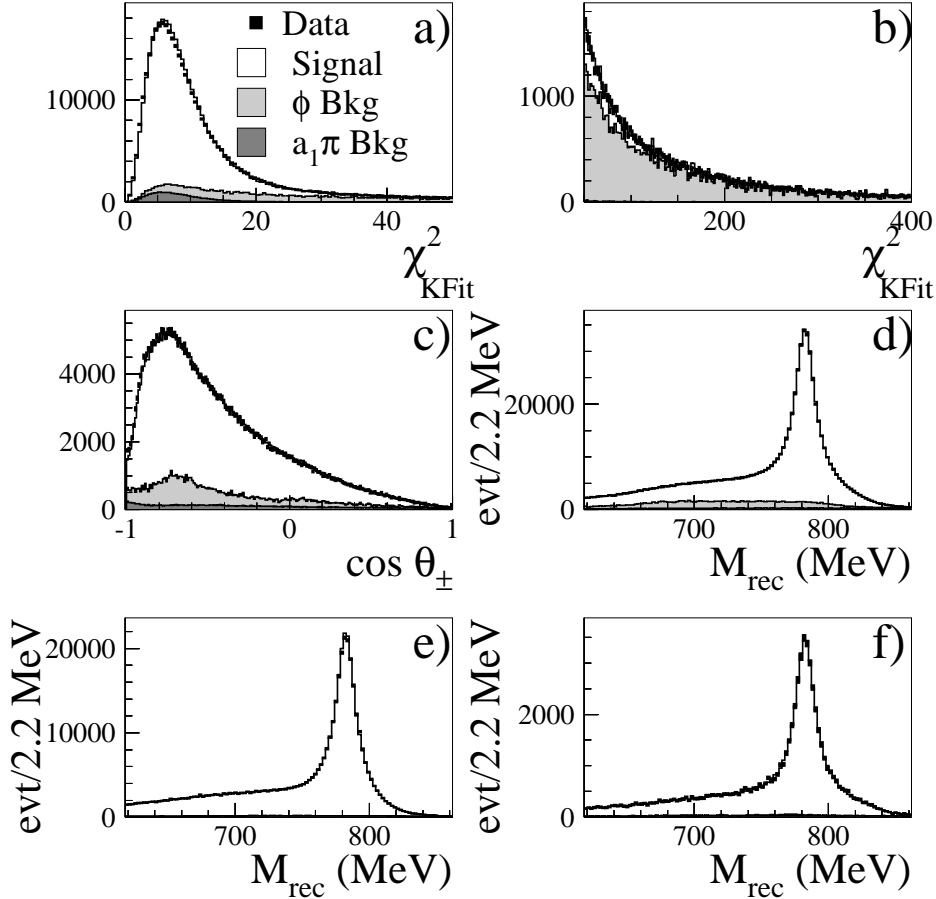}
\end{center}
\caption{
Data-MC comparison for \wpc\ events taken at $\sqrts = 1019.75$ MeV:
$\chi^2$ of kinematic fit for S$_{\rm evts}$ (a) and B$_{\rm evts}$ (b)
samples, angle between charged pions in the $\omega$ rest frame (c) and
\pio\ recoil mass (d) for S$_{\rm evts}$ events. The last distribution
is shown also at $\sqrts = 1000.10$ MeV (e) and $\sqrts = 1029.95$ MeV 
(f). }
\label{Fig:DataMC_wpc}
\end{figure}

The signal counting is performed for each \sqrts\ bin by 
fitting the \pio\ recoil mass ($M_{\rm rec}$) 
distribution for both, S$_{\rm evts}$ and B$_{\rm evts}$ samples with MC 
signal and resonant, non-resonant background shapes.
The fit procedure is based on a likelihood function which takes 
into account both data and MC statistics.
In Fig.~\ref{Fig:DataMC_wpc}.a-d, data-MC comparisons for events in
the most populated bin of \sqrts\ are shown. 
In Fig.~\ref{Fig:DataMC_wpc}.e-f,
the $M_{\rm rec}$ distribution is also shown for the two outermost center of
mass energies. 

The results are summarized in Tab.~\ref{Tab:counting}, where the 
signal counting ($N^{4\pi}$) 
and the visible cross section ($\sigma_{\rm vis}^{4\pi}$) 
are reported for all \sqrts\ bins. Errors on $\sigma_{\rm vis}^{4\pi}$
include statistics, background subtraction and a systematic error of 
0.75\%, dominated by the luminosity measurement (0.5\%). The other
systematics being due to cosmic ray veto, event counting and final 
state radiation.

We have checked the stability of the result by repeating the 
whole analysis chain with a wide variation of the selection criteria on:
(a) minimum cluster energy and angle in the TW definition;
(b) value of $\chi^2_{\rm Kfit}$ cut used to select S$_{\rm evts}$ and
B$_{\rm evts}$ samples;
(c) distribution used for the B$_{\rm evts}$ sample when performing 
signal counting;
(d) data/MC efficiency correction curves. 
The resulting $\sigma_{\rm vis}^{4\pi}$ values are then used to estimate
the systematic error of the measurement. The fit to the visible cross 
section as a function of \sqrts\ described in Sec.~\ref{Par:FitResults} 
is repeated for all of these variations. The quoted systematics 
uncertainty is calculated as the quadratic sum of RMS's obtained from 
variations (a)-(d).

\subsection{$\eeto\omega\pio\to\wpn$}

The selection for $\pio\pio\gamma$ starts requiring five 
neutral clusters in the prompt Time Window with $E_\gamma \ge 7$ MeV 
and polar angle $|\cos\theta_\gamma| < 0.92$. 
After a first kinematic fit (Fit1, $N_{\rm \! dof}=9$) imposing 
total 4-momentum conservation and time of flight, photons are paired to 
\pio's by minimizing a $\chi^2$ function, built using the invariant mass 
of the two $\gamma\gamma$ pairs. A second kinematic fit (Fit2, 
$N_{\rm \! dof}=11$) imposes also the constraint on \pio\ masses.

\begin{figure}[!ht]
\begin{center}
\includegraphics[width=.9\textwidth]{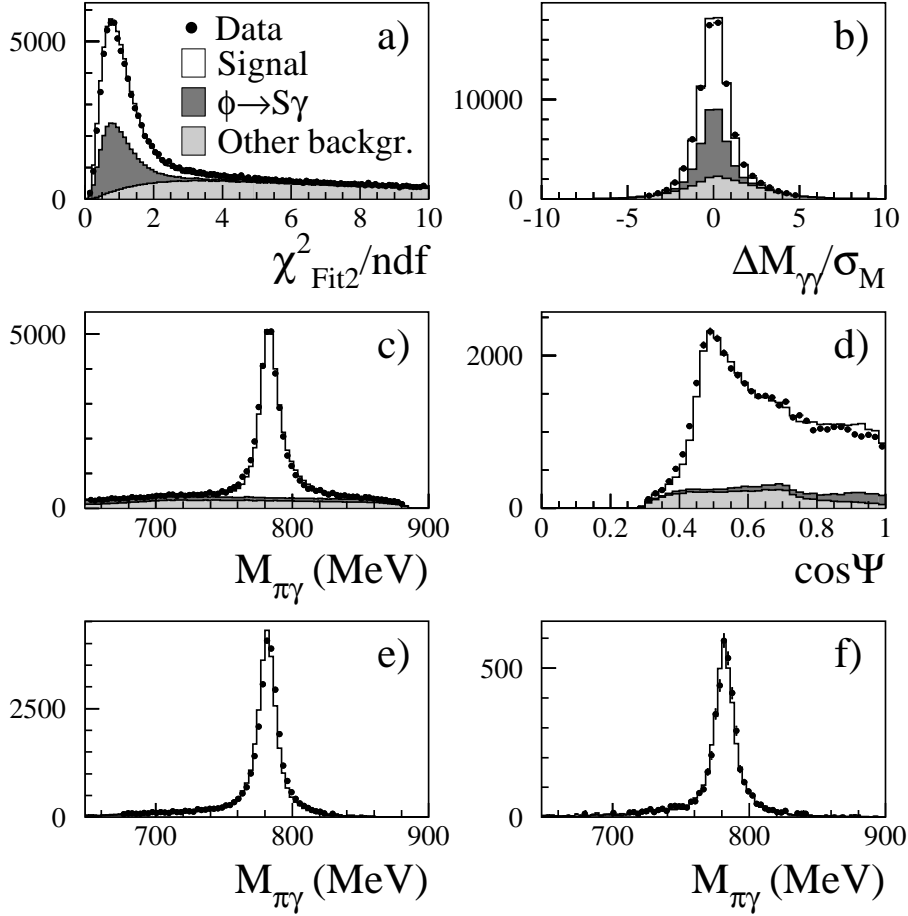}
\end{center}
\caption{
Data-MC comparison for $\pio\pio\gamma$ events taken at 
$\sqrts = 1019.75$ MeV:
(a) normalized $\chi^2$ of the second kinematic fit after 
acceptance cuts (Fit2),
(b) normalized two-photon invariant mass after $\chi^2_{\rm Fit2}$ cut,
(c) $\pio\gamma$ invariant mass after $\Delta M_{\gamma\gamma}$ cut,
(d) $\cos\psi$ distribution after all analysis cuts;
$M_{\pi\gamma}$ distribution at $\sqrts = 1000.10$ MeV 
(e) and $\sqrts = 1029.95$ MeV (f).}
\label{Fig:DataMC_wpn}
\end{figure}

The background is rejected by requiring $\chi^2_{\rm Fit2}/{\rm Ndf}\le 5$ and
$\Delta M_{\gamma\gamma}= 
|M_{\gamma\gamma}-M_{\pi}| \le 5\,\sigma_{\gamma\gamma}$, where 
$M_{\gamma\gamma}$ and $\sigma_{\gamma\gamma}$ are evaluated using 
the photon momenta from Fit1. 
After these cuts the remaining sample is dominated by 
$\epm\to\omega\pio\to\pio\pio\gamma$ and $\phi\to S\gamma\to\pio\pio\gamma$
events. Signal is then selected neglecting the interference 
between these two processes and cutting on the intermediate state mass.
Defining $M_{\pi\gamma}$ as the closest mass to $M_{\omega}$ of the 
two $\pio\gamma$ combinations, only events satisfying the requirement
$750 < M_{\pi\gamma} < 830$ MeV are retained.
The residual background contamination ($\sim 20\%$ at the $\phi$ peak) 
comes predominantly from $\phi\to\eta\gamma\to\pio\pio\pio\gamma$ events, 
where two photons are lost or merged.

In Fig.~\ref{Fig:DataMC_wpn}.a-d, data-MC comparison for events in the 
most populated \sqrts\ bin are shown. The $\psi$ variable is the minimum 
angle between the photon and the \pio's in the di-pion rest frame.
A good agreement is observed both after acceptance selection and after
applying analysis cuts. The comparison for the $M_{\pi\gamma}$ distribution
is also shown for the two outermost center of mass energies, where the
background contribution is practically negligible 
(Fig.\ \ref{Fig:DataMC_wpn}.e-f).

The overall signal selection efficiency is 
evaluated by applying the whole analysis chain to signal MC events: 
$\varepsilon_{\ppg} \sim 46\%$, almost independent from \sqrts.
The value obtained for each bin, together with the corresponding 
integrated luminosity, has been applied to the signal counting to obtain 
the visible cross section ($\sigma_{\rm vis}^{\ppg}$). 
Results are summarized in Tab.\ \ref{Tab:counting}.
Errors on $\sigma_{\rm vis}^{\ppg}$ include statistics, background 
subtraction and a systematic error of 0.6\%, dominated by the luminosity 
measurement (0.5\%). The other systematic effects, related to cosmic 
ray veto and to the event classification filter, have been evaluated with
downscaled data samples.

The stability of the results has been checked by
repeating the fit with: 
(a) a variation of the selection criteria on the TW definition and on 
$\chi^2_{\rm Fit2}$ and $M_{\pi\gamma}$ cuts;
(b) a rescaling of the background components according to the value 
obtained by fitting background-enriched distributions on data with the 
corresponding MC components;
(c) a subtraction to the visible cross section of the interference effect 
between $\phi\to S\gamma$ and the signal. This interference contribution 
is extracted 
from a fit to the Dalitz plot of the $\pio\pio\gamma$ final state 
\cite{Dalitz_piopiog} at the $\phi$ peak, which provides the parameters
describing the two processes. These values are then used to evaluate
for each \sqrts\ the interference contribution, which is subtracted to the 
corresponding $\sigma_{\rm vis}^{\pi\pi\gamma}$. The largest contribution 
is 1.4\%.

\begin{table}[!htb]
  \caption{Signal counting, visible cross section and radiative correction for 
    \eeto\wpc\ and \eeto\wpn\ events. The radiative correction values are
      calculated using for the bare cross section parameters in 
      Tab.~\ref{Tab:fitres}.}
  \label{Tab:counting}
  \newcommand{\m}{\hphantom{$-$}}
  \renewcommand{\tabcolsep}{0.6pc}  
  \renewcommand{\arraystretch}{1.1} 
  \begin{center}
    { \scriptsize
    \begin{tabular}{@{}c|rcc|rcc} \hline\hline
      \sqrts\ (MeV) & 
      \0\0N$^{4\pi}\0\0\0\0$  & $\sigma^{4\pi}_{\rm vis}$ (nb)  & $\delta_{rad}^{4pi}$& 
      \0\0\0N$^{\pi\pi\gamma}\0\0\0$  & $\sigma^{\ppg}_{\rm vis}$ (nb) & $\delta_{rad}^{\ppg}$\\ 
      \hline
       1000.10 & $221917 \pm \0  562$ & $5.72 \pm 0.05$ & 0.885 & $   27110 \pm   167$ &  $0.550 \pm 0.005$ & 0.889 \\
       1009.90 & $ 25968 \pm \0  173$ & $6.20 \pm 0.06$ & 0.893 & $\0  2958 \pm \0 56$ &  $0.581 \pm 0.012$ & 0.895 \\
       1017.20 & $ 16209 \pm \0  171$ & $5.71 \pm 0.08$ & 0.924 & $\0 2108 \pm \0 46$  &  $0.564 \pm 0.018$ & 0.923 \\
       1018.15 & $ 22167 \pm \0  158$ & $5.60 \pm 0.06$ & 0.933 & $\0  2557 \pm \0 60$ &  $0.541 \pm 0.014$ & 0.944 \\
       1019.30 & $  4799 \pm \0\0 90$ & $5.88 \pm 0.12$ & 0.918 & $\0\0 322 \pm \0 22$ &  $0.480 \pm 0.034$ & 0.964 \\
       1019.45 & $ 58077 \pm \0  340$ & $5.89 \pm 0.06$ & 0.912 & $\0  6058 \pm   101$ &  $0.497 \pm 0.009$ & 0.962 \\
       1019.55 & $ 93596 \pm \0  445$ & $5.93 \pm 0.05$ & 0.908 & $\0  9516 \pm   130$ &  $0.487 \pm 0.008$ & 0.960 \\
       1019.65 & $171571 \pm \0  888$ & $5.98 \pm 0.05$ & 0.904 & $   18349 \pm   189$ &  $0.509 \pm 0.007$ & 0.958 \\
       1019.75 & $326774 \pm \0  872$ & $6.04 \pm 0.05$ & 0.900 & $   34049 \pm   282$ &  $0.505 \pm 0.006$ & 0.955 \\
       1019.85 & $256008 \pm    1248$ & $6.08 \pm 0.05$ & 0.896 & $   26124 \pm   234$ &  $0.508 \pm 0.006$ & 0.951 \\
       1019.95 & $ 35850 \pm \0  263$ & $6.20 \pm 0.07$ & 0.893 & $\0  3510 \pm \0 74$ &  $0.491 \pm 0.011$ & 0.948 \\
       1020.05 & $ 17971 \pm \0  167$ & $6.21 \pm 0.08$ & 0.889 & $\0  1843 \pm \0 52$ &  $0.516 \pm 0.016$ & 0.944 \\
       1020.15 & $  8190 \pm \0  132$ & $6.23 \pm 0.11$ & 0.886 & $\0\0 702 \pm \0 32$ &  $0.501 \pm 0.024$ & 0.940 \\
       1020.45 & $  9657 \pm \0  117$ & $6.41 \pm 0.09$ & 0.878 & $\0\0 667 \pm \0 31$ &  $0.488 \pm 0.024$ & 0.928 \\
       1022.30 & $ 16931 \pm \0  141$ & $7.24 \pm 0.08$ & 0.869 & $\0  1891 \pm \0 43$ &  $0.612 \pm 0.018$ & 0.891 \\
       1023.00 & $ 29611 \pm \0  177$ & $7.41 \pm 0.07$ & 0.871 & $\0  3101 \pm \0 61$ &  $0.619 \pm 0.013$ & 0.888 \\
       1029.95 & $ 33681 \pm \0  186$ & $7.84 \pm 0.07$ & 0.887 & $\0  3896 \pm \0 65$ &  $0.689 \pm 0.013$ & 0.892 \\
      \hline \hline      
    \end{tabular}
    } 
  \end{center}
\end{table}

\section{Fit results and $\omega$ branching ratios extraction}
\label{Par:FitResults}

\begin{figure}[!ht]
\begin{center}
\includegraphics[width=1.\textwidth]{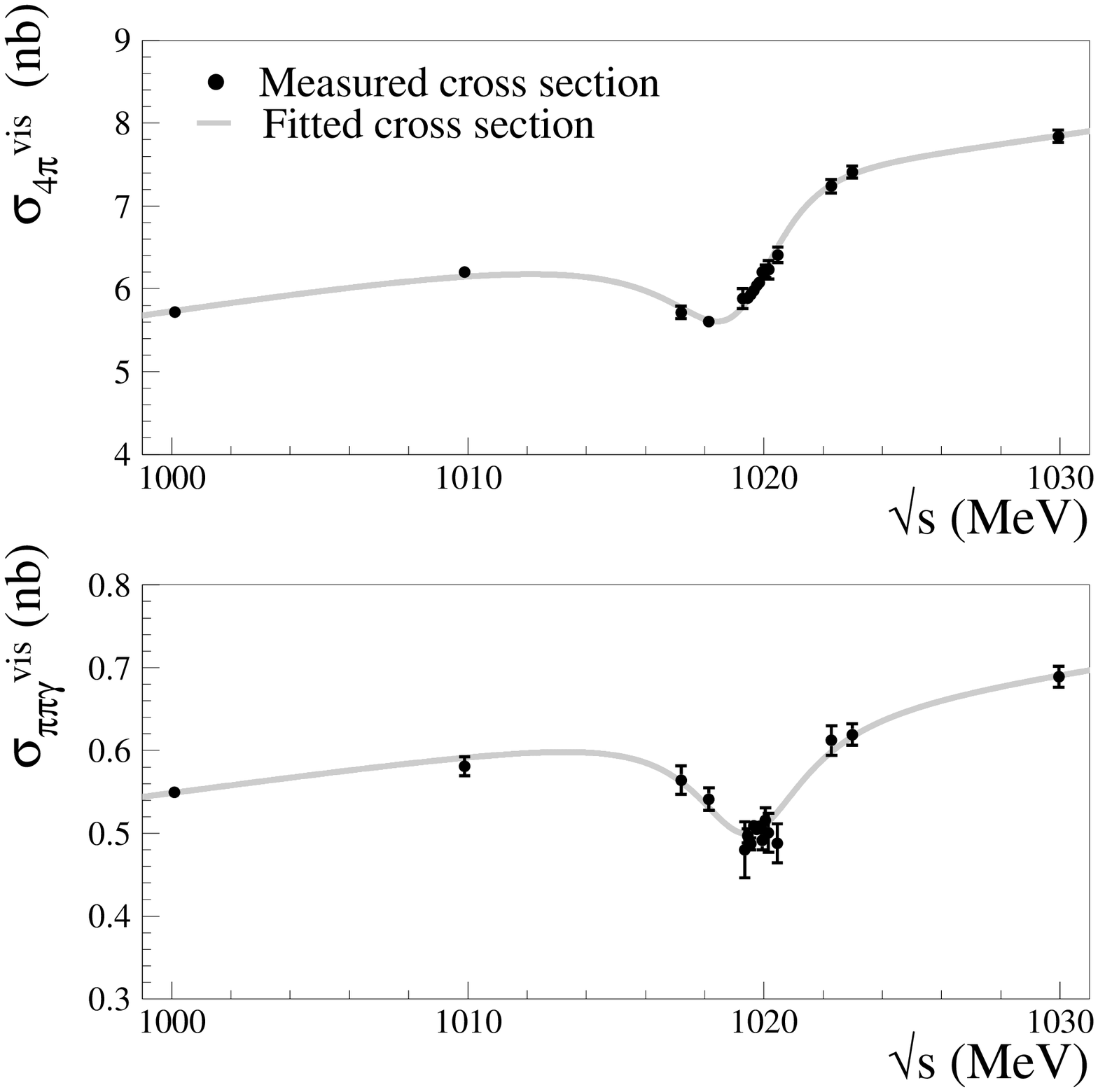}
\end{center}
\caption{Cross section fit results for the \eeto\wpc\ (top) and
\eeto\wpn\ (bottom) channels. 
Black dots are data, solid line is the resulting fit function.}
\label{Fig:cross}
\end{figure}

The measured values of visible cross section, shown in 
Tab.~\ref{Tab:counting}, are fitted with the parametrization 
(\ref{eq:xsec}) convoluted with the radiator function 
\cite{RAD}. The free fit parameters of the bare cross section are: 
$\sigma_{0}^j$, $\Re(Z_j)$, $\Im(Z_j)$ and $\sigma'_j$,
where $j$ represents either the $4\pi$ or $\ppg$ final state.
In Fig.~\ref{Fig:cross} data points with the superimposed fit result 
are shown for both channels. The values of the parameters 
are reported in Tab.~\ref{Tab:fitres}. 
The second error quoted is the systematic uncertainty. It is evaluated as the 
quadratic sum of the RMS of the parameters extracted by repeating the fit 
with different conditions, as described in the previous section.
The resulting $\chi^2/N_{\rm \! dof}$ are 11.79/13 ($P(\chi^2)=54\%$) 
and 4.78/13 ($P(\chi^2)=98\%$) for \wpn\ and \wpc\ channel, respectively.
The fit has been repeated using a VMD model \cite{phiwp2} 
based on $\rho$ and $\rho'$ intermediate states 
for the non-resonant term of the cross section. 
Results are in agreement with the linear parametrization
within one standard deviation.\footnote{ 
The fitting function used for the non-resonant term is:
$$
\sigma_{nr}^j(E) = \sigma_0^{j} 
\frac{m_\phi^3}{E^3}
\frac{\left|m_\rho^2\Pi_\rho(E)+A_j m_{\rho'}^2\Pi_{\rho'}(E)\right|^2}
     {\left|m_\rho^2\Pi_\rho(m_\phi)+ A_j m_{\rho'}^2\Pi_{\rho'}(m_\phi)\right|^2}
\frac{P_{\rm f}(E)}{P_{\rm f}(m_\phi)}
$$
where $\Pi_{\rho(')}(E) = (m_{\rho(')}^2-E^2-iE\Gamma_{\rho(')}(E))^{-1}$ is
the vector meson propagator while $P_{\rm f}(E)$ 
describes the energy dependence 
of the phase space volume of the final state. For the latter we assume the 
approximation of an infinitely narrow $\omega$ meson. 
The free parameters in this case are $\sigma_0^j$ and the real number $A_j$.
This parametrization allows us to compare directly the value of $\sigma_0^j$ 
obtained from this model with the value in Tab.~\ref{Tab:fitres}.
The width of the $\rho$ meson has a dependence with energy 
$\Gamma_\rho(E)=\Gamma_\rho(m_\rho)(m_\rho/E)(p_\pi(E)/p_\pi(m_\rho))$,
while the $\rho'$ width is assumed fixed.
$m_{\rho(')}$ and $\Gamma_{\rho(')}$ values are 
taken from \cite{PDG07}. 
For the parameter A we obtain: $-0.15\pm0.04$, where the error takes 
into account only the contributions from Tab.~\ref{Tab:counting}.
}

\begin{table}[!htb]
  \caption{Fit results for the \eeto\wpc\
    and for \eeto\wpn cross section.} 
  \label{Tab:fitres}
  \newcommand{\m}{\hphantom{$-$}}
  \renewcommand{\tabcolsep}{0.6pc} 
  \renewcommand{\arraystretch}{1.2} 
  \begin{center}
    \begin{tabular}{@{}l|c|c}
      \hline\hline
      Parameter & $\eeto\wpc$ & $\eeto\wpn$ \\\hline
        $\sigma_0$ [nb]   &\m$ 7.89  \pm 0.06\0\pm 0.07 $ &\m$ 0.724  \pm 0.010 \0\pm 0.003 $\\
        $\Re e(Z)$        &\m$ 0.106 \pm 0.007 \pm 0.004$ &\m$ 0.011  \pm 0.015 \0\pm 0.006 $\\
        $\Im m(Z)$        &  $-0.103 \pm 0.004 \pm 0.003$ &  $-0.154  \pm 0.007 \0\pm 0.004 $\\
        $\sigma'$ [nb/MeV]&\m$ 0.064 \pm 0.003 \pm 0.001$ &\m$ 0.0053 \pm 0.0005  \pm 0.0002$\\\hline\hline
      \end{tabular}
  \end{center}
\end{table}

After removing common systematics on the luminosity (0.5\%),
from the two measurements we obtain:
\begin{equation}
  \frac{\sigma_0(\omega\to\pio\gamma)}{\sigma_0(\omega\to\pip\pim\pio)} = 
  0.0918\pm 0.0016
\end{equation}
Taking into account the phase space of the two decays 
\cite{phiwp2}, the ratio of the partial widths is:
\begin{equation}
  \frac{\Gamma(\omega\to\pio\gamma)}{\Gamma(\omega\to\pip\pim\pio)} = 
  0.0897\pm 0.0016
\end{equation}

\begin{figure}[!ht]
  \begin{center}
    \includegraphics[width=0.7\textwidth]{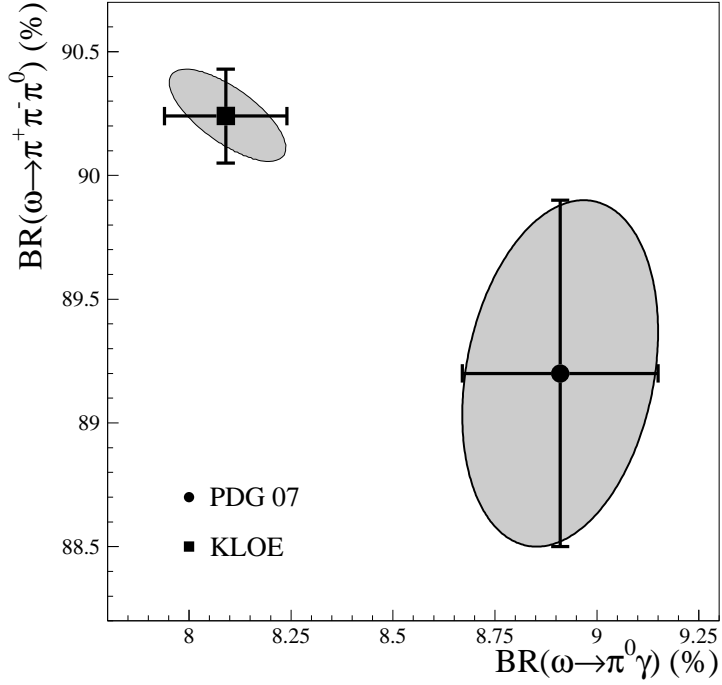}
  \end{center}
  \caption{
    Branching fraction for the two main $\omega$ decay channels. 
    The black square is the KLOE fit result, while the black dot is the
    constrained fit result in \cite{PDG07}. The gray ellipses are the 68\% 
    C.L.\ regions.}
  \label{Fig:wbrfit}
\end{figure}

Since these two final states account for $98\%$ of the $\omega$ 
total width, we use the 
$\Gamma(\omega\to\pio\gamma)/\Gamma(\omega\to\pip\pim\pio)$ 
ratio and the sum of rarer BR's \cite{PDG07} to obtain from a fit:
\begin{eqnarray}
  \br(\omega\to\pip\pim\pio) & = & (90.24\pm 0.19)\% \\
  \br(\omega\to\pio\gamma)\ \ \ \ \ \,    & = & (\08.09\pm 0.14)\%
\end{eqnarray}
with a correlation of -71\%. Comparison between our evaluation and the 
values in \cite{PDG07} is shown in Fig.\ \ref{Fig:wbrfit}. 
Our result for $\br(\omega\to\pio\gamma)$ is less than the PDG value by three 
standard deviations. It is in good agreement with the recent prediction in 
\cite{dipion}.

\section{BR($\phi\to\omega\pio$) evaluation}

The measured $\sigma_0^{4\pi}$ and $Z_{4\pi}$ parameters of the \wpc\ final 
state are related to the BR(\f\to\wpo) through the relation:
\begin{equation}
\br(\phi\to\wpo) = \sigma_0(m_{\phi})|Z_{4\pi}|^2\frac{1}{\sigma_{\phi}}
= \frac{\sigma_0^{4\pi}|Z_{4\pi}|^2}{BR(\omega\to\pip\pim\pio)}
\frac{M_{\f}^2}{12\pi\Gamma(\f\to\ep\el)},
\end{equation}
where $\sigma_0(m_{\phi})$ is the total cross section of the $\eeto\wpo$ 
process and $\sigma_{\phi}$ is the peak value of the bare cross 
section for the \f\ resonance. 
Using the parameters obtained from the \wpc\ analysis, the $\Gamma_{ee}$ 
measurement from KLOE \cite{GeeKLOE} for the evaluation of $\sigma_{\phi}$,
and our value for  $\br(\omega\to\pip\pim\pio)$ we extract:
\begin{equation}
  \br(\phi\to\wpo) = (4.4\pm 0.6)\times 10^{-5}
\end{equation}

The error is reduced by a factor of two with respect to the best previous 
measurement from the SND experiment \cite{phiwp2}, which is in agreement 
with our result.

\section{Conclusions}

Using a sample of 600 pb$^{-1}$ collected at center of mass energy 
between 1000 and 1030 MeV, we have measured the cross section 
parameters for the two processes \epm\to\wpc\ and \epm\to\wpn,
obtaining the ratio 
$\Gamma(\omega\to\pio\gamma)/\Gamma(\omega\to\pip\pim\pio)$ 
with an accuracy of 1.8\%.
This ratio, together with the unitarity relation and the  
BR measurements on the other $\omega$ decay channels, substantially
improves the accuracy on the dominant $\omega$ branching fractions 
giving a value for BR($\omega\to\pio\gamma$) which is  
three standard deviations lower than the Particle Data Group fit 
\cite{PDG07}.
Moreover, the parameters describing the \eeto\wpc\ reaction around 
$M_\phi$ are used to extract the most precise BR measurement of 
the OZI and G-parity violating decay \f\to\wpo.

\section*{Acknowledgements}

We thank the DAFNE team for their efforts in maintaining low background running 
conditions and their collaboration during all data-taking. 
We want to thank our technical staff: 
G.F.Fortugno and F.Sborzacchi for their dedicated work to ensure an
efficient operation of 
the KLOE Computing Center; 
M.Anelli for his continuous support to the gas system and the safety of
the
detector; 
A.Balla, M.Gatta, G.Corradi and G.Papalino for the maintenance of the
electronics;
M.Santoni, G.Paoluzzi and R.Rosellini for the general support to the
detector; 
C.Piscitelli for his help during major maintenance periods.
This work was supported in part
by EURODAPHNE, contract FMRX-CT98-0169; 
by the German Federal Ministry of Education and Research (BMBF) contract 06-KA-957; 
by the German Research Foundation (DFG),'Emmy Noether Programme',
contracts DE839/1-4;
and by the EU Integrated
Infrastructure
Initiative HadronPhysics Project under contract number
RII3-CT-2004-506078.

%

\end{document}